\documentclass[a4paper]{jpconf}

\usepackage{amsmath,amsfonts,amssymb}
\usepackage{mathrsfs}
\usepackage{graphicx}
\usepackage[small]{caption}
\usepackage[utf8]{inputenc}

\usepackage{color}

\begin{document}
\title{One--loop corrections for Higgs--portal dark matter}

\author{J.~Armando Arroyo \& Sa\'ul Ramos-S\'anchez}

\address{Instituto de F\'isica, Universidad Nacional Aut\'onoma de M\'exico,
POB 20-364, Cd.Mx. 01000, M\'exico}

\ead{armandoarroyo@fisica.unam.mx, ramos@fisica.unam.mx}

\begin{abstract}
Models endowed with Higgs portals can probe into the hidden sectors of particle physics
while providing stable dark matter candidates. Previous tree--level
computations in such scenarios have shown that experimental bounds
constrain dark matter to a very narrow region in parameter space.
Aiming at improving the study of the implications of those constraints, 
we inspect one--loop corrections to the annihilation cross section for scalar 
dark matter into observable fermions. 
We find that these loop contributions might be enough to drastically change
those results by deforming in about 10\% the allowed parameter space 
for dark matter particles with masses even below 1 TeV. These findings
encourage further investigation.
\end{abstract}

%%%%%%%%%%%%%%%%%%%%%%%%%%%%%%%%%%%%%%%%%%%%%%%%%%%%%%%%%%%%%%%%%%%%%%%%%%%%%%%%%%%%%%%%%%%%%%%%%%%%%%%%%%%%
% Intro
%%%%%%%%%%%%%%%%%%%%%%%%%%%%%%%%%%%%%%%%%%%%%%%%%%%%%%%%%%%%%%%%%%%%%%%%%%%%%%%%%%%%%%%%%%%%%%%%%%%%%%%%%%%%

\section{Introduction}
Since the early days of cosmology~\cite{Zwicky:1933xx}, the quest for
an understanding of dark matter (DM) and its origin has led to several proposals. 
These include massive astrophysical compact halo objects (MACHOs)~\cite{Bird:2016dcv}, 
weakly interacting massive particles (WIMPs)~\cite{Bertone:2004pz}
and axion-like particles (ALPs)~\cite{Sikivie:2006ni}. WIMPs have been considered as optimal 
DM candidates mainly because of their natural appearance in many models that could explain other
longstanding issues of particle physics. Their theoretical appeal has triggered an exhaustive experimental search
by, among others, the XENON100 and LUX collaborations, that has resulted in severe constraints 
on the WIMP parameter space~\cite{Aprile:2012nq,Akerib:2015rjg}.
Verifying that existing models comply with these strict bounds requires now that all 
their predicted observables, such as the production and decay rates,  
be computed with maximal precision.

On the other hand, the recent confirmation of the existence of the Higgs particle
with a relatively large mass opens up a new set of possibilities. Especially, 
WIMPs might well couple directly to the Higgs field,
establishing a connection known as {\it a Higgs portal}~\cite{Patt:2006fw} 
between the {\it dark sector} and the Standard Model (SM).
If there were such couplings, WIMPs would e.g. decay into visible matter via 
the exchange of a Higgs boson, affecting direct and indirect detection signals. 

Beyond DM, Higgs portals may be relevant to address other intriguing questions, 
such as the nature of cosmological inflation~\cite{Gross:2015bea}, the instability in the 
electroweak vacuum~\cite{Falkowski:2015iwa}, and more recently, the 750 GeV diphoton 
excess reported at the LHC \cite{Choi:2016cic}. Despite their great potential, in
this work we shall only focus on the features of DM in these models.

The additional particles required in Higgs portals are considered to be {\it hidden}
in the sense that they do not carry SM charges. They are dynamical (real) 
scalars $S$, spinors $\psi$ or vectors $V_\mu$, confined to interact with the 
SM via Lagrangian couplings with the Higgs field $H$, such as
\begin{equation}
S^2 H^\dagger H\,,\qquad
\bar{\psi} \psi H^\dagger H\,,\quad
\quad V_\mu V^\mu H^\dagger H\,,
\end{equation}
and affected by their own (in principle, arbitrary) potential energy density.
The scalar case clearly requires the inclusion of a $\mathbb Z_2$ symmetry 
(or another similar mechanism) to avoid tadpole contributions and to ensure DM stability. 
These different particles are good DM candidates as long as they comply with experimental bounds.
In this respect, fermions are disfavored because they do not only yield a non--renormalizable model, but
they are already excluded by XENON data when considering only thermal DM~\cite{Djouadi:2011aa}. 

The phenomenological implications of Higgs portals with scalars and vectors as DM 
have been studied in~\cite{Djouadi:2011aa, Lebedev:2011iq} (and a recent global analysis for 
the scalar case has been done in~\cite{Han:2015hda}), where only tree--level
contributions to the decay and production rates have been considered. It was shown
in those works that experimental bounds of WMAP~\cite{2011ApJS..192...18K} and XENON100~\cite{Aprile:2012nq}  
allow for a very reduced set of values in the parameter space of the models,
so that radiative corrections may become important to falsify them. Although
one--loop corrections to DM--nucleon scattering have been already analyzed~\cite{Hisano:2015bma},
the corresponding study for DM annihilation processes is still missing. To motivate and
to exhibit the importance of completing this task is one of the aims of the present work.

Based on the results obtained in~\cite{Arroyo:2015xx}, we discuss here the main aspects of
Higgs portals and explore the importance of one--loop corrections to the cross sections of
the processes associated with DM detection.
The structure of this contribution is as follows. In section~\ref{sec:portals}, we give a brief review 
of the scalar and vector Higgs portals; in section~\ref{sec:corrections} we present the one--loop
analysis for the cross section in the context of the scalar portal done with the aid of FeynArts~\cite{Hahn:2000kx} 
and FormCalc~\cite{Hahn:1998yk} and finally, in section~\ref{sec:Discussion} we discuss the 
obtained results and ongoing work.

%%%%%%%%%%%%%%%%%%%%%%%%%%%%%%%%%%%%%%%%%%%%%%%%%%%%%%%%%%%%%%%%%%%%%%%%%%%%%%%%%%%%%%%%%%%%%%%%%%%%%%%%%%%%
% Higgs portals
%%%%%%%%%%%%%%%%%%%%%%%%%%%%%%%%%%%%%%%%%%%%%%%%%%%%%%%%%%%%%%%%%%%%%%%%%%%%%%%%%%%%%%%%%%%%%%%%%%%%%%%%%%%%
\section{Scalar and vector Higgs portals}
\label{sec:portals}

The scalar and vector Higgs portals have been studied previously in~\cite{Djouadi:2011aa, Lebedev:2011iq} 
and here we will only revise the main features.

The simplest example of these models is the scalar portal. It is achieved by introducing
a (real) massive scalar field $S$ subject to a $\mathbb Z_2$ symmetry, under which $S$ is odd and 
all SM fields are even. The full renormalizable Lagrangian then reads
\begin{equation} 
\mathcal{L} ~=~ \mathcal{L}_{SM} + \frac12 \partial_\mu S\,\partial ^\mu S 
             -  \frac{1}{2}\mu_S^2 S^2 - \frac{\lambda_{hs}}{2}S^2H^\dagger H - \frac{\lambda_S}{4} S^4\,,
\end{equation}
where $\lambda_{hs}$, $\mu_S$ and $\lambda_S$ are the portal and the two self--interaction real parameters. 
After electroweak symmetry breaking (EWSB), the Higgs--portal potential becomes
\begin{equation} 
\label{scalarV}
 V(h,S) ~=~ \frac12 m_S^2 S^2 + \frac{\lambda_{hs}v}{2} S^2\,h  + \frac{\lambda_{hs}}{4} S^2\,h^2 + \frac{\lambda_S}{4} S^4\,, 
\end{equation}
where
$m_S^2 = \mu_S^2 + \lambda_{hs}v^2/{2}$
is the physical mass of $S$, $v\approx 246\, GeV$ denotes the electroweak vacuum expectation value (VEV)
of the Higgs field $H$, such that $H(x)\to (v + h(x))/\sqrt2$ (and $S$ has vanishing VEV) in the vacuum, 
with the small Higgs field perturbation $h$.

Similarly, for the vector Higgs portal we introduce a vector field $V_\mu$, probably
stemming from a U(1) gauge symmetry, whose Higgs--portal potential is given by
\begin{equation} 
\label{eq:vectorV}
\mathcal{L} \supset \frac{1}{2}\mu_V^2 V_\mu V^\mu + \frac{\lambda_{hv}}{2}V_\mu V^\mu H^\dagger H + \frac{\lambda_V}{4}\left(V_\mu V^\mu\right)^2\,.
\end{equation}
Since the mass term $V_\mu V^\mu$ is not gauge invariant if $V_\mu$ arises from a gauge symmetry,\footnote{
$V_\mu$ may have a different origin. E.g. it could be a composite field, whose effective potential is eq.~\eqref{eq:vectorV}.} 
some gauge--restoring mechanism, such as the Stueckelberg mechanism,\footnote{For a review on the Stueckelberg mechanism, see \cite{Zhang:2008xq}} 
must be invoked. After EWSB, the physical mass of the vector is given by
$m_V^2 = \mu_V^2+{\lambda_{hv}v^2}/{2}$.

Not only does EWSB generate the physical masses $m_S$ and $m_V$ of the DM particles, but, as
eq.~\eqref{scalarV} evidences, it also ``produces'' the cubic interactions $h\,S^2$ and $h\,V_\mu V^\mu$. 
These interaction terms allow for the possibility of the {\it invisible} decay of the Higgs particle into DM. 
LHC bounds~\cite{PhysRevLett.114.191803} on the associated decay width $\Gamma(h\to SS)$ do not 
constrain DM with masses over few hundred GeV~\cite{Han:2015hda}; hence, this decay
shall not be investigated in this work.  

\begin{figure}[!t!]
\centering
  \includegraphics[scale=0.6, trim={4.2cm 18cm 0cm 4cm},clip]{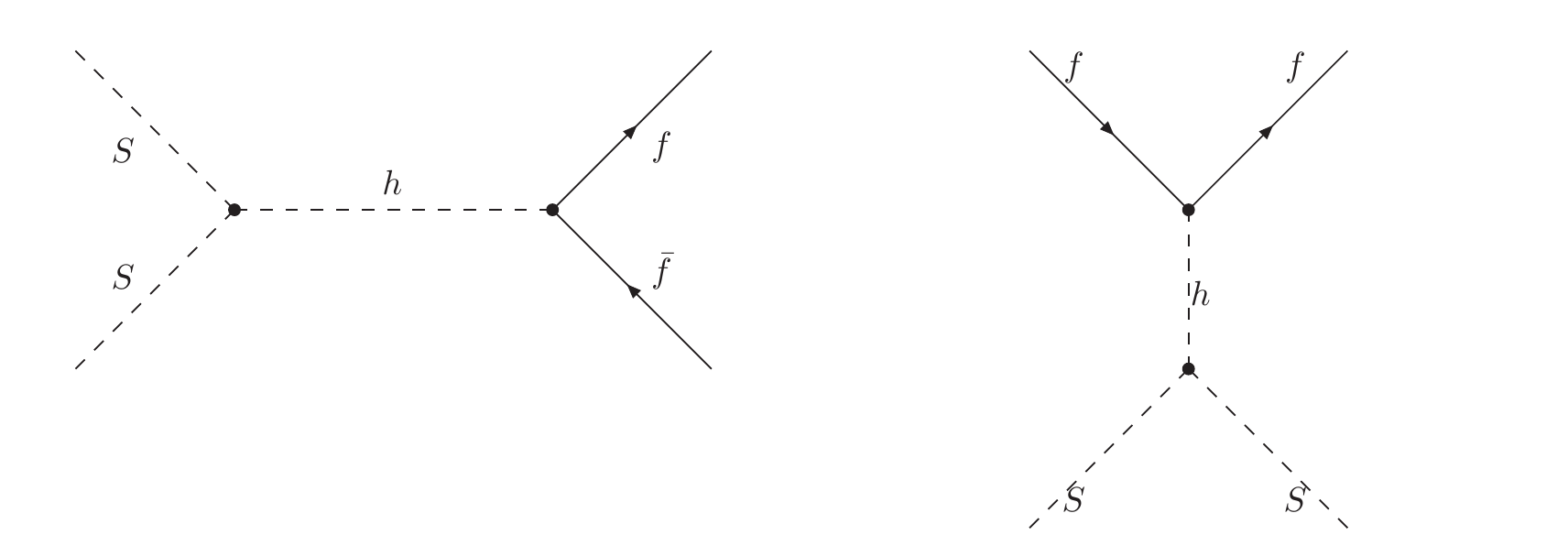}

  \includegraphics[scale=0.6, trim={4.2cm 18cm 0cm 4cm},clip]{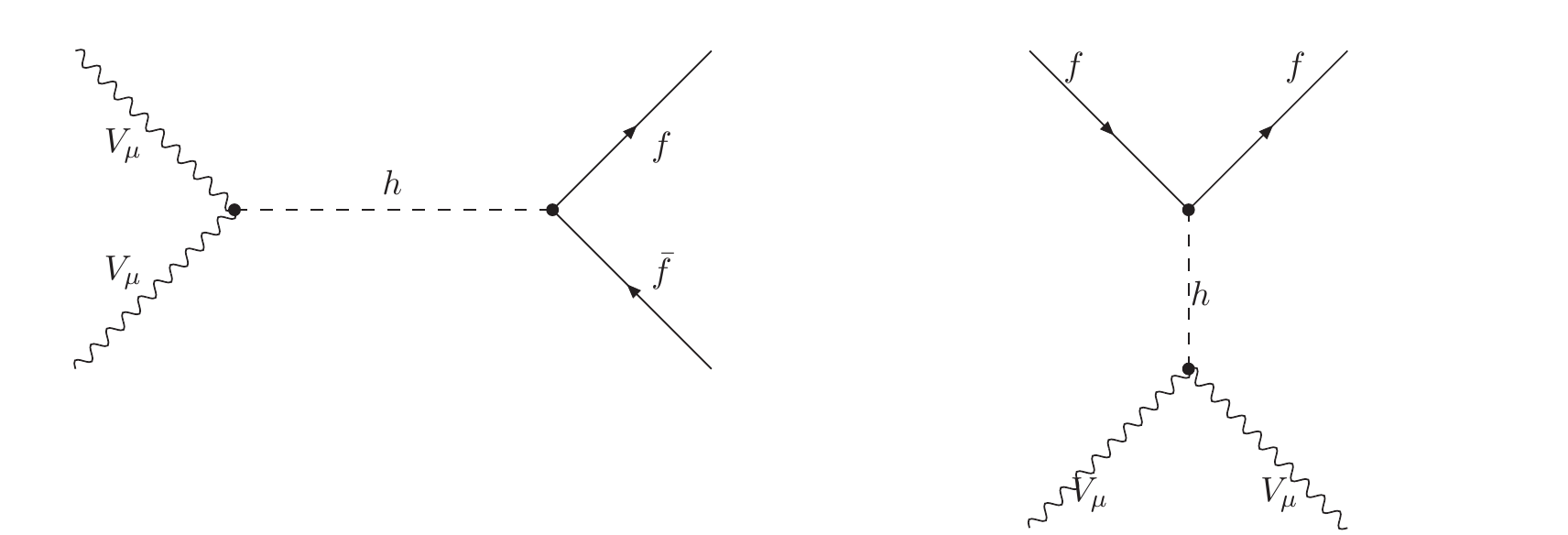}  
  \caption{The left-hand-side diagrams depict DM annihilation into SM fermions, while those on the right,
           DM--fermion scattering for the scalar and vector cases. \label{fig:SS->ff}}
\end{figure}

Additionally, as depicted in Fig.~\ref{fig:SS->ff}, the $hSS$ vertex allows for DM annihilation 
into SM fermions as well as DM--fermion scattering. 
The thermally--averaged cross section of the first process, 
$\langle\sigma_{DM} v_{rel}\rangle$ (with ${}_{DM}$ referring either to $S$ or $V_\mu$), relates to Planck's measurement of DM relic
abundance~\cite{Ade:2015xua}, $\Omega_{CDM}h^2=0.1198\pm0.0015$, roughly by
\begin{equation}
\label{eq:relicconstraints}
\Omega_{CDM} h^2 \sim 3\cdot10^{-27} cm^3\,s^{-1}/{\langle\sigma_{DM} v_{rel}\rangle}\,,
\end{equation}
where $v_{rel}$ denotes the DM relative velocity or M{\o}ller velocity.
On the other hand, DM--fermion scattering is important for direct DM detection,
because the spin--independent DM--nucleon scattering cross section $\sigma_{SI}$ can be 
computed from it under certain assumptions, allowing for a comparison with LUX and XENON100 data.

For scalar and vector Higgs portals, the newest version of micrOMEGAs~\cite{Belanger:2014vza} 
performs the numerical computation of $\sigma_{SI}$ up to one--loop level and compares it
with the LUX and XENON100 bounds. Therefore,
only the computation of radiative corrections to $\langle\sigma_{DM} v_{rel}\rangle$
may be relevant if those contributions are sizable, as we shall show that they turn out to be.

At tree--level, the thermally--averaged annihilation cross sections are given by
\begin{eqnarray}
\label{eq:sigmaS} 
\langle\sigma_S v_{rel}\rangle &=& \frac{\lambda_{hs}^2m_f^2}{16\pi}\frac{(1-m_f^2/m_S^2)^{3/2}}{(4m_S^2-m_h^2)^2}\,,\\
\label{eq:sigmaV}
\langle\sigma_V v_{rel}\rangle &=& \frac{\lambda_{hv}^2m_f^2}{48\pi}\frac{(1-m_f^2/m_V^2)^{3/2}}{(4m_V^2-m_h^2)^2}\,,
\end{eqnarray}
for the scalar and vector case, respectively. In these eqs., $m_f$ is the
mass of the fermions in the final state. Note that the cross section eq.~\eqref{eq:sigmaS} 
(eq.~\eqref{eq:sigmaV}) depends solely on $\lambda_{hs}\,(\lambda_{hv})$ and $m_S\,(m_V)$. 
It then follows that Planck's data bound on $\Omega_{CDM}$ leads to constraints on these parameters.

%%%%%%%%%%%%%%%%%%%%%%%%%%%%%%%%%%%%%%%%%%%%%%%%%%%%%%%%%%%%%%%%%%%%%%%%%%%%%%%%%%%%%%%%%%%%%%%%%%%%%%%%%%%%
% 1-loop corrections
%%%%%%%%%%%%%%%%%%%%%%%%%%%%%%%%%%%%%%%%%%%%%%%%%%%%%%%%%%%%%%%%%%%%%%%%%%%%%%%%%%%%%%%%%%%%%%%%%%%%%%%%%%%%

\section{One--loop corrections in the scalar portal}
\label{sec:corrections}
Sizable one--loop contributions to DM cross sections in Higgs portals may become important as they may 
close further the narrow region of allowed parameters, thereby finding the model already falsified by data. 

\begin{figure}[ht]
\centering
  \includegraphics[scale=0.45]{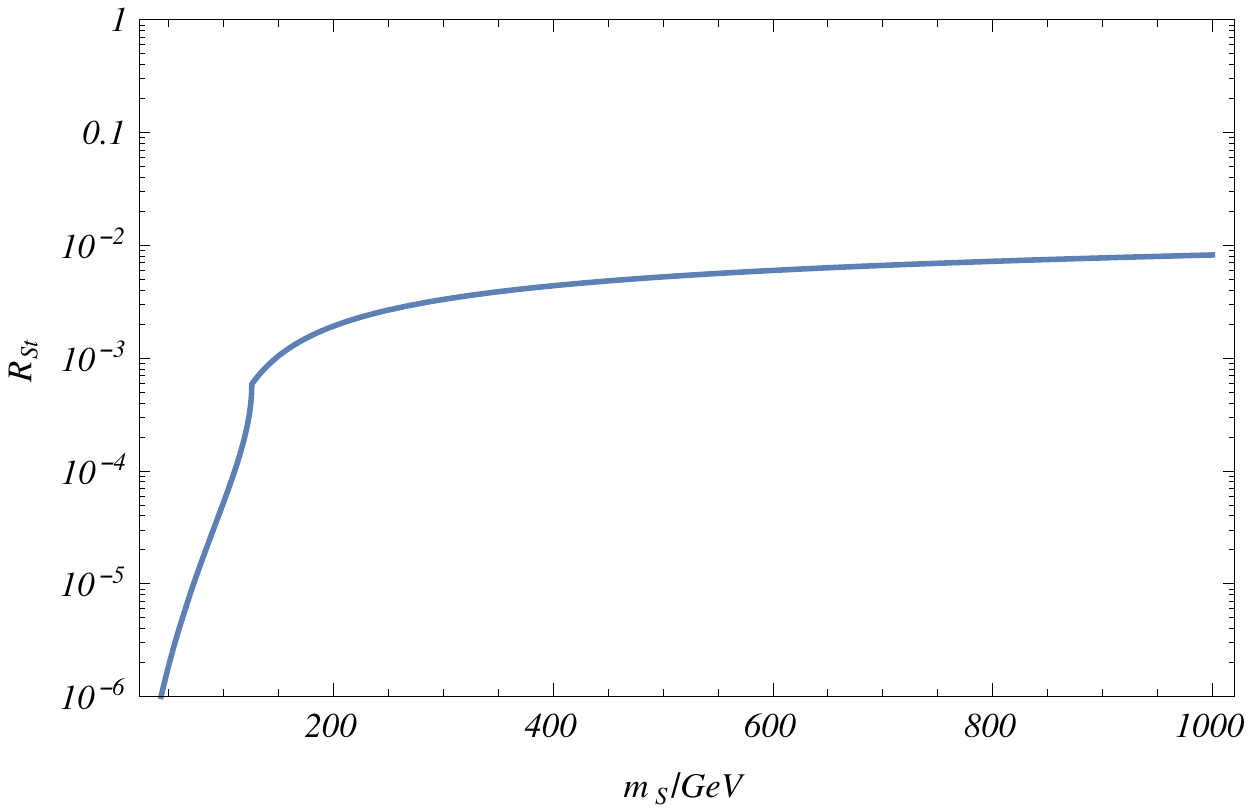}
~
  \includegraphics[scale=0.45]{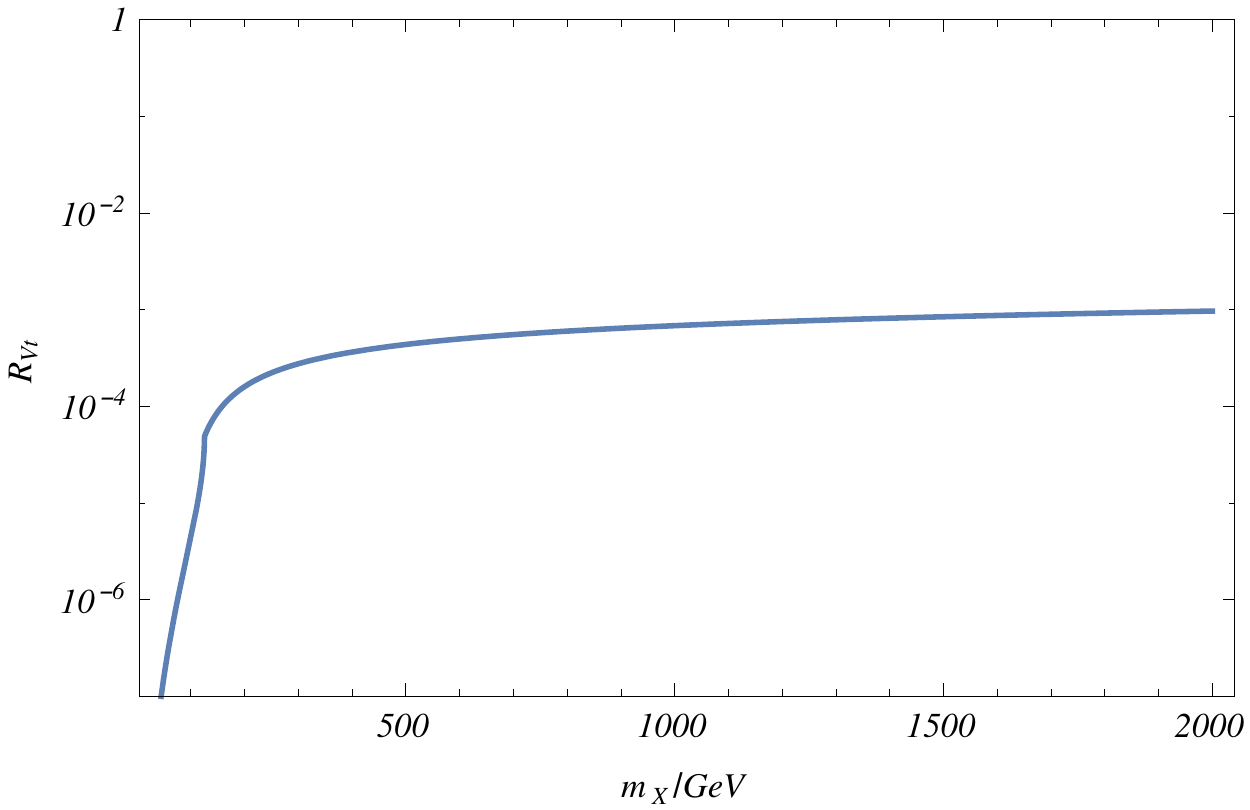}
\caption{Approximate comparison between the amplitudes at one--loop and at tree--level 
for the process $SS\to t\bar{t}$ (left) and $V_\mu V^\mu\to t\bar{t}$ (right).\label{fig:ratio}}
\end{figure}

In a previous analysis~\cite{Arroyo:2015xx}, we studied the WIMP mass evolution of the ratio between the simplest 
one--loop amplitude correction and the tree--level amplitude for DM annihilation into $t$ quarks,
\begin{equation}\label{eq:ratio}
  R_{DMt}\equiv \frac{|\mathcal{M}_{\text{only 1-loop}}(DMDM\to t\bar{t})|^2}{|\mathcal{M}_{\text{tree}}(DM\,DM\to t\bar{t})|^2}\,,
\end{equation}
where, as before, $DM$ stands for either $S$ or $V_\mu$ WIMP candidates.
This ratio lets us estimate how subdominant one--loop corrections are. As it is evident from Fig.~\ref{fig:ratio} (left),
in the scalar case this ratio shows that, for DM masses around 1 TeV and larger, the simplest one--loop 
contribution analyzed can be as large as $\sim 1\%$ of the tree--level amplitude. This preliminary result 
encourages further study, since other one--loop contributions may enlarge this already sizable correction.
In the vectorial case, this approach reveals (see Fig.~\ref{fig:ratio} (right)) that one--loop contributions
may not be so relevant. Thus, we can safely focus on the corrections to the scalar Higgs portal.

\begin{figure}[b!]
\centering
  \includegraphics[scale=0.5]{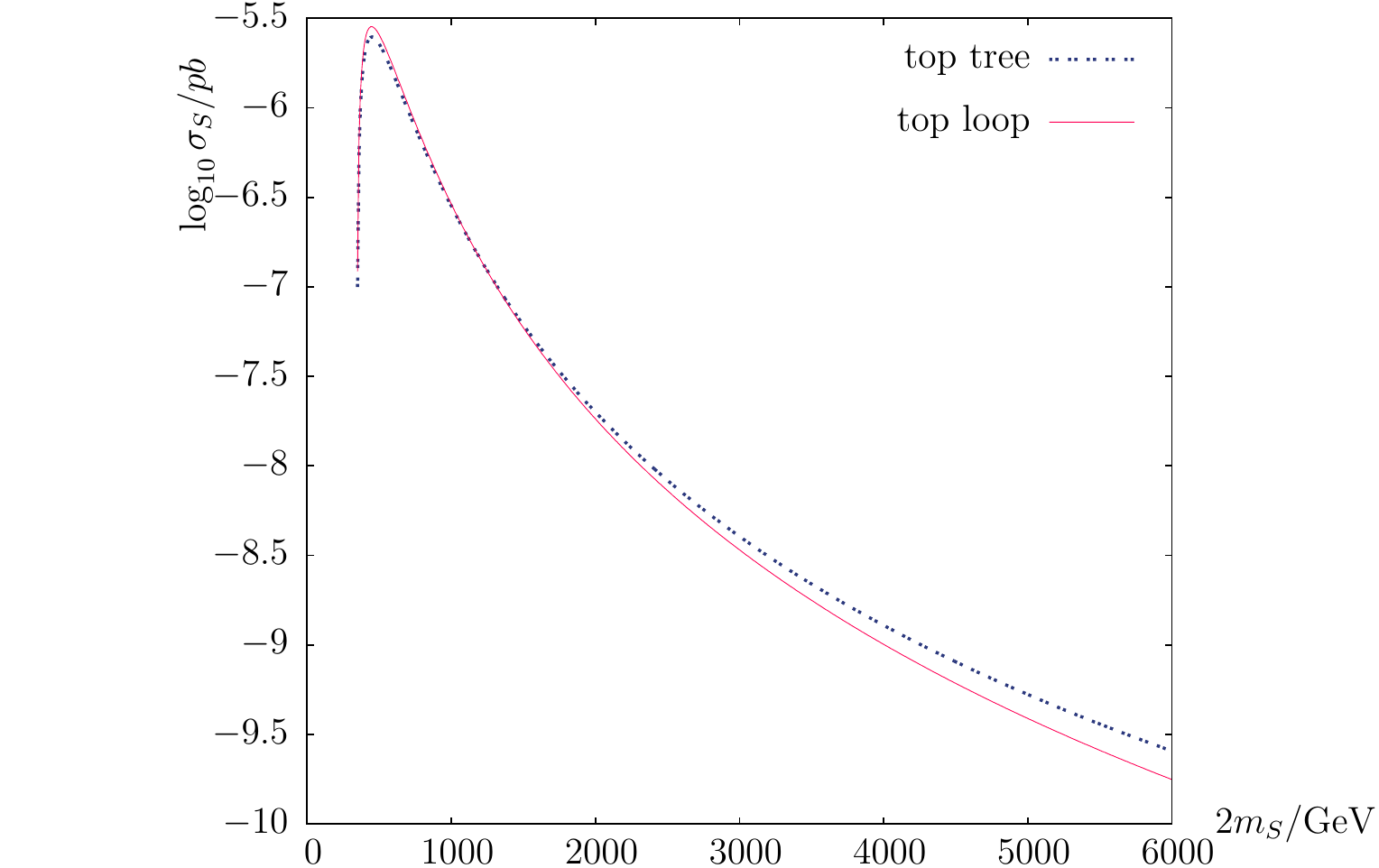}
  \includegraphics[scale=0.5]{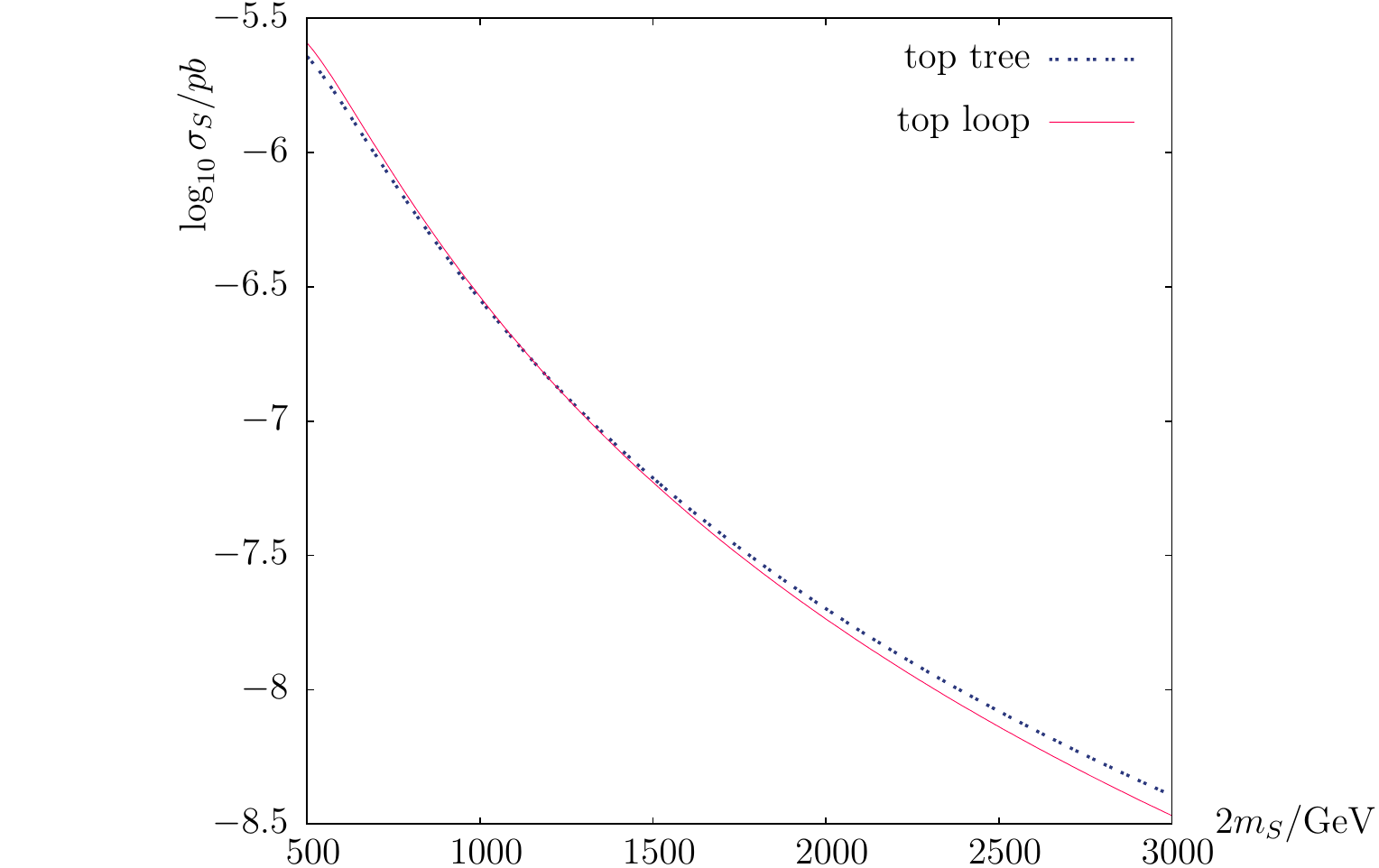}
\caption{Annihilation cross section of scalar DM into $t$ quarks for $\lambda_{hs}=10^{-3}$
         and $m_S$ between $0.25$ and $3$ TeV and a zoomed--in interval. We compare the tree--level (blue/dashed line)
         contribution vs. the full up--to--one--loop cross sections (red/solid line). For $m_S\gtrsim600$ GeV, 
         the one--loop contribution becomes negative and can be (for $m_S\sim 3$ TeV) as large as 30\%  
         of the tree--level cross section.
         \label{fig:annihilationtop}}
\end{figure}

We have extended that previous analysis in two ways. First, we completed the computation of the one--loop
cross section $\sigma_S$ for annihilation into $t$'s, including all possible contributions. Secondly, we
further studied the complete cross section at one--loop level for the annihilation of scalar DM 
into $t$, $b$ and $c$ quarks and $\mu$ and $\tau$ leptons.
The Feynman rules were generated with LanHEP~\cite{Semenov:2008jy}; the 
computation of amplitudes, one loop integrals and the cross sections have been performed by
using the public numerical and analytical tools FeynArts \cite{Hahn:2000kx}, LoopTools and 
FormCalc \cite{Hahn:1998yk}. 

\begin{figure}[h!]
\centering
  \includegraphics[scale=0.5]{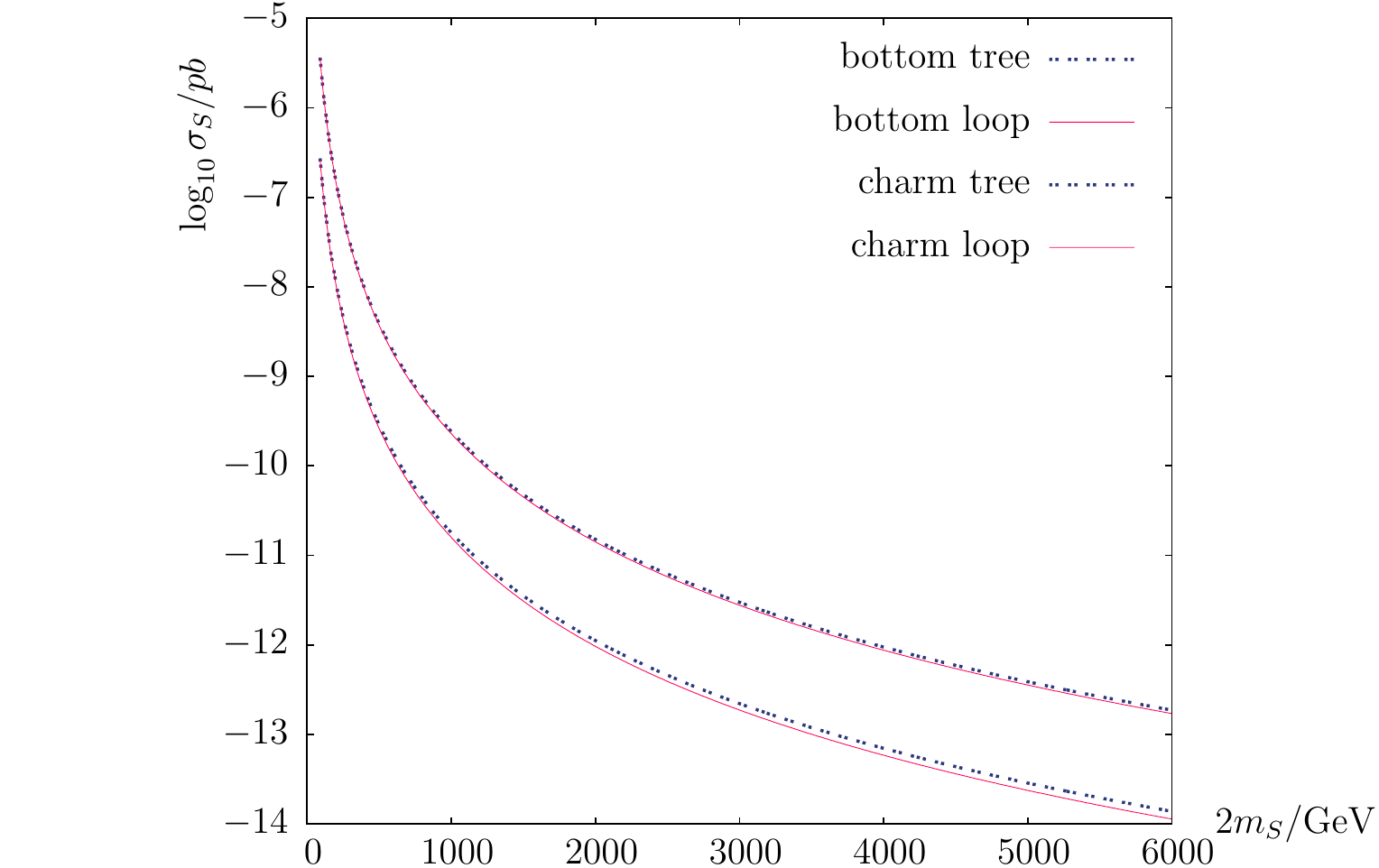}
  \includegraphics[scale=0.5]{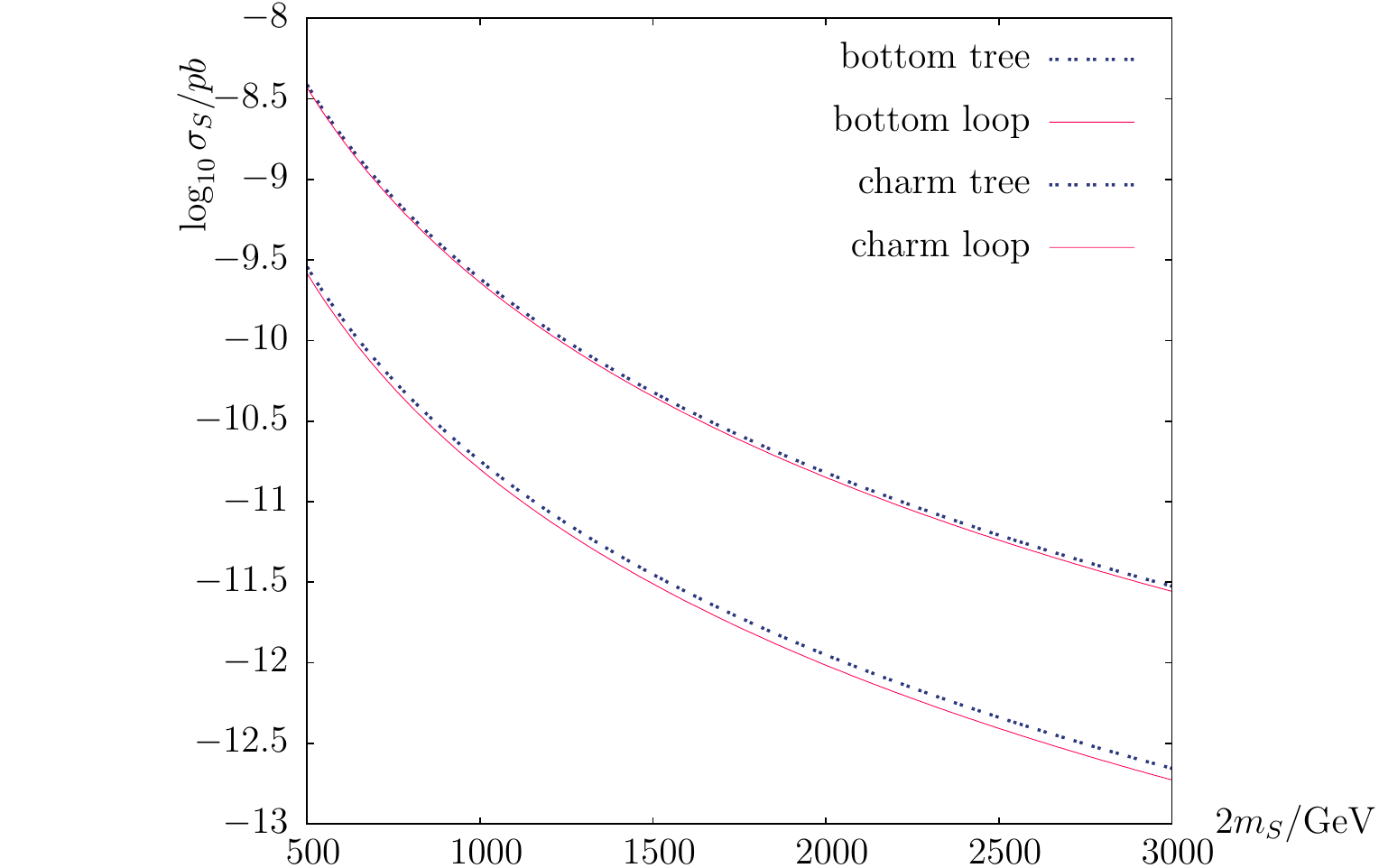}
  
  \includegraphics[scale=0.5]{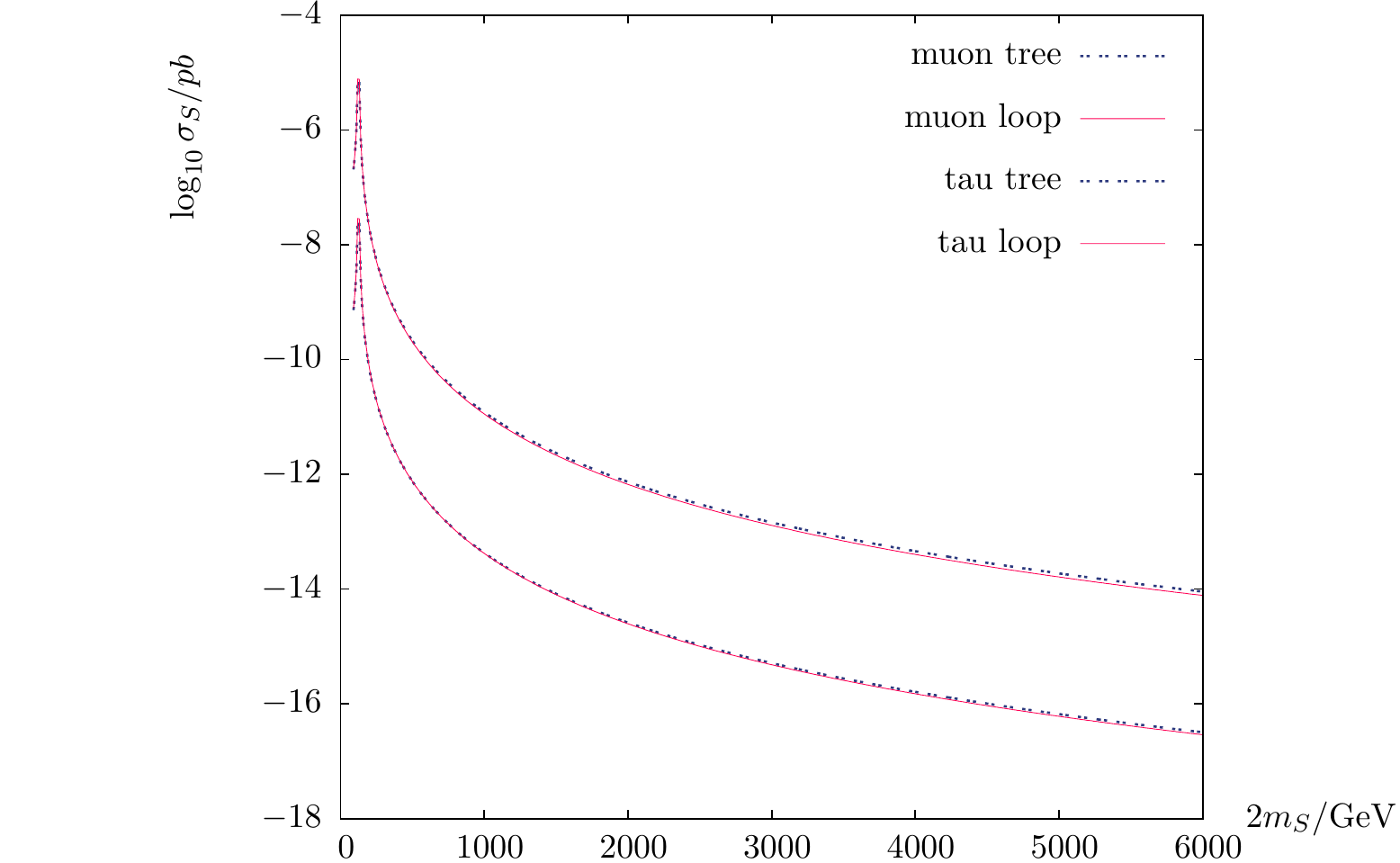}
  \includegraphics[scale=0.5]{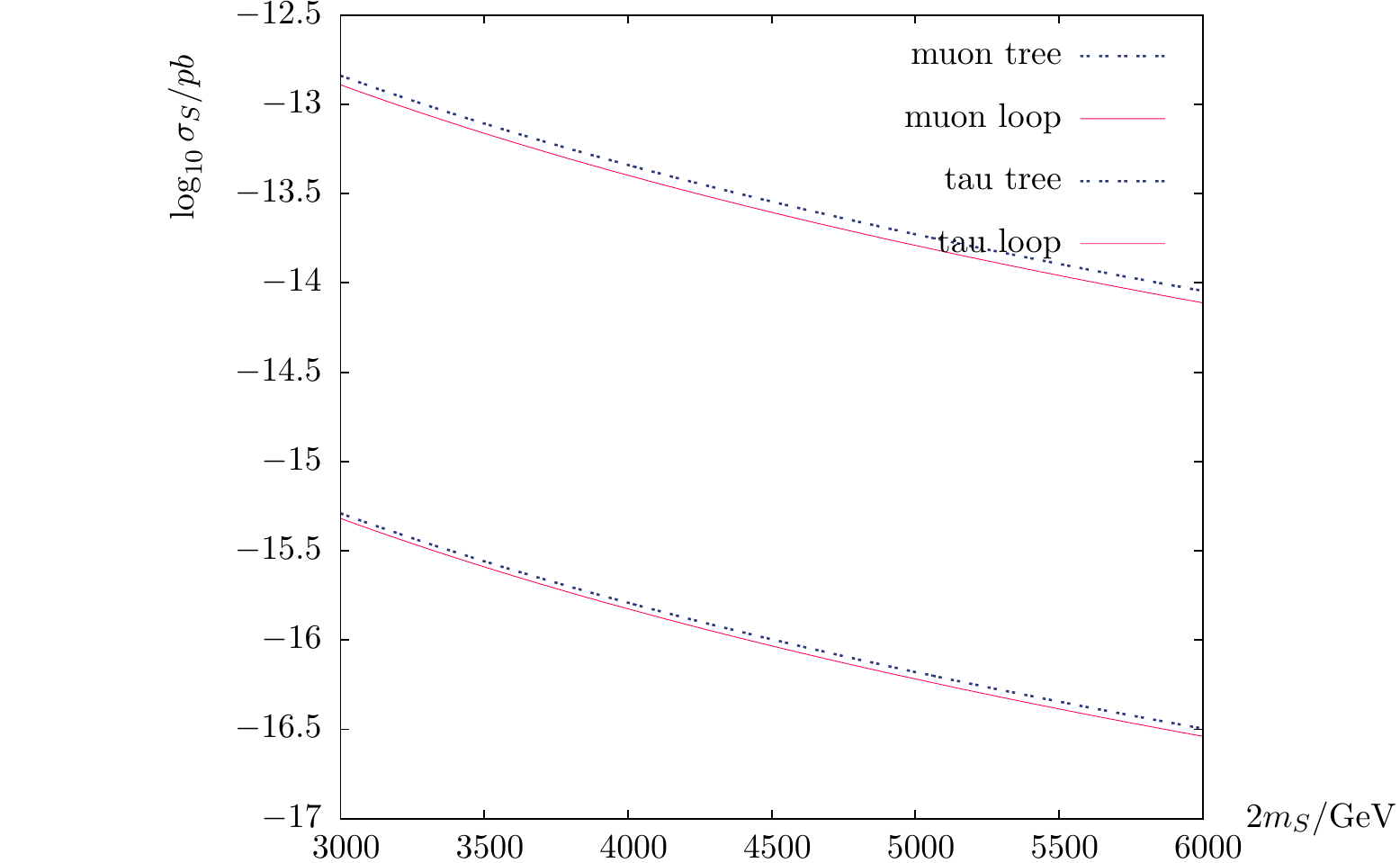}
\caption{\small Annihilation cross section of scalar DM for $\lambda_{hs}=10^{-3}$ and $m_S$ between $45$ GeV and $3$ TeV
         (left panels) and a zoomed--in interval (right panels).
         The upper panels depict the tree--level (blue/dashed line) and the full up--to--one--loop (red/solid line) contributions with
         $b$ and $c$ quarks in the final states, whereas the lower panels display the same contributions with
         $\mu$ and $\tau$ leptons in the final states.
         \label{fig:annihilation}}
\end{figure}

Our results are displayed in Figs.~\ref{fig:annihilationtop} and~\ref{fig:annihilation}. Fig.~\ref{fig:annihilationtop}
shows the comparison of the tree--level $\sigma_{S,\text{tree}}$ and full up--to--one--loop $\sigma_{S,\text{1-loop}}$ 
cross sections for scalar DM annihilation into $t$ quarks and $\lambda_{hs}=10^{-3}$. We note that for $2m_S\gtrsim1.2$ TeV,  
this contribution is negative and significantly larger than $1\%$, becoming as large as $20\%$ for
scalar Higgs--portal WIMPs with $m_S\sim 2$ TeV. A similar comparison
has been done for DM annihilation into  light quarks and heavy leptons (Fig.~\ref{fig:annihilation}), to confirm our
expectation that top contributions are dominant. For different values of $\lambda_{hs}$, we find similar results.

%%%%%%%%%%%%%%%%%%%%%%%%%%%%%%%%%%%%%%%%%%%%%%%%%%%%%%%%%%%%%%%%%%%%%%%%%%%%%%%%%%%%%%%%%%%%%%%%%%%%%%%%%%%%
% Final discussion
%%%%%%%%%%%%%%%%%%%%%%%%%%%%%%%%%%%%%%%%%%%%%%%%%%%%%%%%%%%%%%%%%%%%%%%%%%%%%%%%%%%%%%%%%%%%%%%%%%%%%%%%%%%%

\section{Punchline and outlook}
\label{sec:Discussion}

We have computed the one--loop corrections to the annihilation cross section of Higgs--portal DM
into SM fermions. Our preliminary results stress the need to carefully determine the influence 
of radiative corrections in the computation of the cross sections in models with Higgs portals.
We notice that, particularly for the scalar Higgs portal and certain parameter values, the one--loop
contributions can be negative and as large as $13\%$ ($21\%$) of the tree level cross section for 
admissible DM with masses of about 1 TeV (2 TeV).

The resulting thermally averaged cross sections $\langle \sigma_S v_{rel}\rangle$ are functions of the two parameters
of the Higgs portal, $m_S$ and $\lambda_{hs}$, such that for each value of $m_S$ an interval of $\lambda_{hs}$
is still allowed by Planck's data, according to eq.~\eqref{eq:relicconstraints}. Since the computed
cross section $\sigma_S$ is reduced by a sizable amount for heavy WIMPs, 
so is $\langle \sigma_S v_{rel}\rangle$. Thus, the predicted relic abundance increases, stressing
the tension between Higgs portals and measurements by WMAP, LUX and XENON100, when the
one--loop corrections to the DM--nucleon scattering amplitude $\sigma_{SI}$ are additionally included.
The detailed report of this analysis shall be presented elsewhere~\cite{Arroyo:2016xxx}.

%%%%%%%%%%%%%%%%%%%%%%%%%%%%%%%%%%%%%%%%%%%%%%%%%%%%%%%%%%%%%%%%%%%%%%%%%%%%%%%%%%%%%%%%%%%%%%%%%%%%%%%%%%%%
% Thank you, guys
%%%%%%%%%%%%%%%%%%%%%%%%%%%%%%%%%%%%%%%%%%%%%%%%%%%%%%%%%%%%%%%%%%%%%%%%%%%%%%%%%%%%%%%%%%%%%%%%%%%%%%%%%%%%
\section*{Acknowledgments}

We are grateful for useful discussions with T. Hahn and A. Ibarra.
This work was partly supported by CONACyT grant 151234.
S.~R-S. would like to thank the ICTP for the kind hospitality and support received
through its Junior Associateship Scheme during the realization of this work.

%%%%%%%%%%%%%%%%%%%%%%%%%%%%%%%%%%%%%%%%%%%%%%%%%%%%%%%%%%%%%%%%%%%%%%%%%%%%%%%%%%%%%%%%%%%%%%%%%%%%%%%%%%%%
% Bibliography
%%%%%%%%%%%%%%%%%%%%%%%%%%%%%%%%%%%%%%%%%%%%%%%%%%%%%%%%%%%%%%%%%%%%%%%%%%%%%%%%%%%%%%%%%%%%%%%%%%%%%%%%%%%%

\section*{References}
%\bibliography{bibliography}
%\bibliographystyle{iopart-num}

\providecommand{\newblock}{}

\end{document}